\documentclass[twocolumn,pra,aps,superscriptaddress, twocolumn]{revtex4}

\usepackage[utf8]{inputenc}
\usepackage[english]{babel}
\usepackage[T1]{fontenc}
\usepackage{amsmath}

\usepackage{mathrsfs} % to use mathscr font
\usepackage{xspace,amsmath,amsfonts,amsthm,amssymb,amsbsy}
\usepackage{graphicx}
\usepackage{amssymb}
\usepackage{color}
\usepackage{psfrag}
\usepackage{ifsym}
\usepackage[usenames,dvipsnames]{xcolor}
\usepackage[colorlinks=true,citecolor=Blue,linkcolor=Purple,urlcolor=Cerulean]{hyperref}
\usepackage{epstopdf}
\usepackage{units} % for nicefrac
\usepackage{bbm}
\usepackage{bbold}
\usepackage{ragged2e}

\graphicspath{{../images/}}

\definecolor{pink}{rgb}{1,0.078,0.57}

\definecolor{green}{rgb}{0,0.7,0.9}

\newcommand{\ket}[2] {| #1 \rangle_{#2}}
\newcommand{\bra}[2] {\langle #1 |_{#2}}

\newcommand{\bp}{\mathbf{p}}

\newcommand{\dg}{^{\dagger}}

\newcommand{\br}{\mathbf{r}}
\newcommand{\bR}{\mathbf{R}}

\newcommand{\Tr}{\mathrm{Tr}}

\newcommand{\therm}[1]{\rho{_{\mathrm{th}}^{#1}}}
\newcommand{\vac}{\ket{\mathrm{vac}}{}}
\newcommand{\cav}{\bra{\mathrm{vac}}{}}

\begin{document}

\title{Thermal States and Wave Packets}

\author{Aur\'elia Chenu}
\email[]{achenu@mit.edu}
\affiliation{Massachusetts Institute of Technology, Cambridge, Massachusetts 02139, USA}
\author{Agata M. Bra\'nczyk}
\affiliation{Perimeter Institute for Theoretical Physics, Waterloo, Ontario, N2L 2Y5, Canada}
\author{J. E. Sipe}
\affiliation{Department of Physics, 60 Saint George Street, University of Toronto, Toronto, Ontario, M5R 3C3 Canada}

\begin{abstract}
The classical and quantum representations of thermal equilibrium are strikingly different, even for free, non-interacting particles. While the first involves particles with well-defined positions and momenta, the second usually involves energy eigenstates that are delocalized over a confining volume. In this paper, we derive convex decompositions of the density operator for non-interacting, non-relativistic particles in thermal equilibrium that allow for a connection between these two descriptions. Associated with each element of the decomposition of the $N$-particle thermal state is an $N$-body wave function, described as a set of wave packets; 
the distribution of the average positions and momenta of the wave packets can be linked to the classical description of thermal equilibrium, while the different amplitudes in the wave function capture the statistics relevant for fermions or bosons.
\end{abstract}

\maketitle

Thermal states are ubiquitous in
physics. In both classical and quantum statistical mechanics they
play a central role in many calculations, either as an assumption
of the initial state of a system of interest before a perturbation
is applied, or as a description of the state of a reservoir interacting
with a system of interest. In the simple textbook problem of the equilibrium
state of non-interacting, non-relativistic particles confined to a
box and at a given temperature, the classical results for thermodynamic
quantities, such as the free energy, are easily recovered as the appropriate
limits of the quantum results. But even under circumstances where
a classical description should suffice, the classical and quantum
descriptions of equilibrium are strikingly different. In classical
statistical mechanics a distribution function is introduced, assigning
well-defined momenta and positions to particles with appropriate probabilities.
In quantum statistical mechanics, the usual representation of
the thermal density operator is in terms of the energy eigenstates, states 
that are extended throughout the confining volume, with the position uncertainty
in each single-particle state thus given by the size of the box itself.
This is unchanging even as the temperature is increased, and the classical
limit should be recovered.

\begin{figure}[t!]
\includegraphics[width=1\columnwidth]{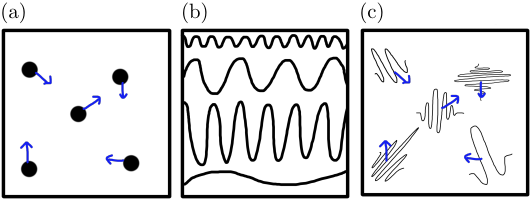}
\caption{
Illustration of various possible representations of thermal equilibrium for free, non-interacting $N$-particles. The classical picture (a) is strikingly different from the quantum picture (b). In the former, each particle has a well-defined position and momentum (blue arrows), while the latter is in terms of eigenstates that spread over the full confining volume. The sketch in (b) depicts a sample of eigenfunctions $\phi_n(\br)$ (Eq. \ref{eq:eigenvalues}) describing the one-particle thermal state (Eq. \ref{eq:rho_conv}). In the main text, we derive alternative decompositions involving wave packets with average positions and momenta. The sketch in (c) depicts the $N$-body wave function, involving  a set of wave packets (Eqs. \ref{eq:wp} and \ref{eq:phi_gen}), and corresponding to a term in this alternative decomposition. The distribution of the average positions and momenta of the wave packets can be linked to the classical description of thermal equilibrium, such that (c) provides a connection between (a) and (b). 
 \label{fig:sketch}}
\end{figure}

Since the early days of quantum mechanics, significant theoretical attention has been put into representing quantum mechanics using the language of classical physics (see, e.g., Schleich \cite{Schleich2011a}). 
Among the first developments is the formalism developed by Moyal \cite{Moyal1949a}, which closely relates  quantum operators and classical functions of phase space variables first suggested by Weyl \cite{Weyl1927a}.  This is based on the observation that the dynamical evolution of the Wigner function, given by the Liouville equation, takes a form identical to the evolution of the density matrix in the classical limit generalizing the Poisson brackets to Moyal brackets. 
Alternative approaches include phase space methods connected either to the corpuscular classical limit---coordinate-momentum phase space (see e.g. Hillery \cite{Hillery1984a} \emph{et al.} for a review)---or to the wave classical limit. These rely on the coherent state representation of bosonic systems \cite{Gardiner2004a,Walls2007a}, which have largely influenced the concepts in quantum optics \cite{Fano1983a} and  have since found many applications.  It was recently shown how the coordinate-momentum and the bosonic coherent state phase-space representations can be identically mapped to each other under the change of variables \cite{Polkovnikov2010a}.

The phase space representation of quantum systems is not unique. 
Focusing on the description of an ideal gas, we here present an alternative representation in coordinate-momentum space that involves a many-body wave function with classical-like variables, and connects the quantum and classical representations of thermal equilibrium. Such a microscopic description of the Gibbs ensemble can assist in deriving thermodynamical quantities from microscopic theory, a question that has attracted attention over the years \cite{Balian2007a} and has remained a subject of debate since the foundations were laid (see e.g. \cite{DAlessio2016a} for a review). In a broader context, and despite known analogies between quantum and classical systems \cite{Strocchi1966a,Frank1994a, Alzar2002a, Novotny2010a,Elze2012a,Briggs2012a}, no general  approach has been successful in establishing a classical representation for statistical mixtures. 

Thermal states are statistical mixtures, and so are usually represented using the density matrix formalism \cite{Fano1957a}. Yet a density matrix generally allows many representations, or `convex decompositions'. \footnote{Specifically, the decomposition of a density matrix $\rho = \sum_\psi p(\psi) \ket{\psi}{} \bra{\psi}{}$ is convex if it satisfies $p(\psi) \geq 0$ for all $\psi$ and $\sum_\psi p(\psi) = 1$.}.
In this paper we construct, for a canonical ensemble describing non-interacting, non-relativistic
particles, convex decompositions of the density operator that involve
sets of wave packets, each wave packet with a localized coordinate
representation and an expectation value of momentum. 
The first decomposition we give  is in terms of static wave-packets.
While this is straightforward for the one-particle subspace, we generalize this decomposition to a many-body formalism using field theory. 
More importantly, we derive an alternative basis that involves dynamic wave-packets and provide a full class of new convex decompositions. 
 We can then see explicitly how the  classical particle picture arises as a limit of the quantum wavepackets;  see Fig. (\ref{fig:sketch}) for a sketched illustration.
The classical limit for the partition function that follows has been studied previously (see, e.g., Huang \cite{HuangBook}). 
Our approach goes beyond these treatments in that we consider the quantum state itself rather than its thermodynamic properties. 

Given the importance
of thermal states in physics, the study of different convex decompositions
of the canonical ensemble is an interesting problem in itself. There
are also practical applications. Earlier \cite{Hornberger2003a}, a
wave packet decomposition of the Maxwell-Boltzmann limit helped in
the study of decoherence of a heavy quantum Brownian particle interacting
with a background gas. 
By representing the ideal gas as an ensemble of quantum particles, its effect on the coherence of the Brownian particle could be obtained from calculating a single scattering event, and averaging over all events to account for the thermal mixture. This theory provided the framework for understanding quantum decoherence experiments with large particles such as $C_{70}$ fullerene \cite{Hornberger2004a, Hackermuller2004a}.
But in \cite{Hornberger2003a,Hornberger2004a, Hackermuller2004a} the gas particles were explicitly
treated as distinguishable; here we employ field theory to treat the
equilibrium of non-interacting fermions or bosons, and construct convex
decompositions of the density operator even at low temperatures where
indistinguishability issues arise. Of course, in the classical limit
the Maxwell-Boltzmann results arise naturally as a special case of
the calculation we present in this paper, but the extensions to fermions
and bosons presented here should allow for more general studies of
decoherence.

The outline of the paper is as follows. 
In Sec. \ref{sec:ensemble} we present a field theory form of the density operator describing the canonical ensemble for free particles, and indicate the standard quantum-mechanical representation.  In Sec. \ref{sec:method} we present the general method used here to derive an alternative convex decomposition, which is applied for a one-particle ensemble in Sec. \ref{sec:N1} and generalized for any number of particles in Sec. \ref{sec:N}. We illustrate how our new decomposition brings insights in the calculation of thermodynamics quantities, specifically the partition function and correlation functions, in Sec. \ref{sec:illu}. Conclusions and connections to previous work, and a generalization to the grand canonical ensemble, are presented in the final section. 

\section{The canonical ensemble}\label{sec:ensemble}
We consider non-interacting,
non-relativistic bosons or fermions of mass $m$ confined by a  square potential
energy $v(\mathbf{r})$ 
 that vanishes everywhere expect on the edges, taken at a large distance $L$ away from the origin, 
thus defining a ``nominal confining box.'' Neglecting spin degrees of freedom, which could be easily
included, the Hamiltonian is given by 
\begin{equation}\label{eq:H}
H=\int\psi^{\dagger}(\mathbf{r})\left(-\frac{\hbar^{2}}{2m}\nabla^{2}+v(\mathbf{r})\right)\psi(\mathbf{r})d\mathbf{r},
\end{equation}
where the operators $\psi\dg(\br)$ and $\psi(\br)$ respectively create and annihilate a particle at position $\br$ and fulfill the commutation relations $\left[ \psi(\br), \psi\dg(\br')\right]_{\pm} = \delta(\br - \br')$ and $\left[ \psi(\br) , \psi(\br')\right]_{\pm} = 0$; here and below the upper of the two signs always refers to fermions and the lower to bosons, and $[A,B]_{\pm} \equiv A B  \pm B A $. Most discussions of thermal equilibrium of such systems
using field theory employ a grand canonical ensemble, where the number
of particles can fluctuate. We turn to this later in the paper, but
mainly consider the canonical ensemble, where the number of particles
is fixed. The relevant subspace $\mathcal{H}^{(N)}$ of $N$ particles
is identified by the identity operator acting over that subspace,
\begin{align}\label{eq:identity}
\begin{split}
\mathbbm{1}^{(N)}\equiv{}& \frac{1}{N!} \int d\br_1 \dots d\br_N \\
\times&  \psi\dg(\br_1) \dots \psi\dg(\br_N) \vac\cav \psi(\br_N) \dots\psi(\br_1) ,
\end{split}
\end{align} 
where $\vac$ indicates the vacuum state with no
particles; this expression is valid for either fermions or bosons. The
density operator for a canonical ensemble involving $N$ particles
can then be constructed by replacing the Hamiltonian (\ref{eq:H}) in the usual Boltzmann factor $\exp(-\beta H)$ by its
projection onto $\mathcal{H}^{(N)}$, i.e. $H\rightarrow\mathbbm{1}^{(N)}H\mathbbm{1}^{(N)}$.
Since the full Hamiltonian (\ref{eq:H}) does not change
the number of particles, it commutes with $\mathbbm{1}^{(N)}$, and
we can write the density operator for a canonical ensemble of $N$
particles at temperature $T$ in a particular form that will prove
convenient, 
\begin{equation} \label{eq:rhoth_def}
\therm{(N)} = \frac{1}{Z^{(N)}} \:   e^{-\beta H/2} \mathbbm{1}^{(N)}\:  e^{-\beta H/2},
\end{equation}
where $Z^{(N)}$ is the partition function for $N$ particles, $\beta=1/k_{B}T$, and $k_{B}$ is Boltzmann's constant; the expression $Z^{(N)}=\Tr[\mathbbm{1}^{(N)}e^{-\beta H}]$
follows from the normalization condition $\Tr[\rho_{\rm th}^{(N)}]=1.$

A starting point for our work below will be
the density operator for a single particle, $\rho_{\rm th}^{(1)}$. The
usual convex decomposition is obtained from the equations above by
expanding the field operator in terms of the energy eigenstates, $\psi(\mathbf{r})=\sum_{n}a_{n}\phi_{n}(\mathbf{r})$,
where $[a_{n},a_{n'}^{\dagger}]_{\pm}=\delta_{nn'}$ and
\begin{equation}\label{eq:eigenvalues}
\left(-\frac{\hbar^{2}}{2m}\nabla^{2}+v(\mathbf{r})\right)\phi_{n}(\mathbf{r})=E_{n}\phi_{n}(\mathbf{r}).
\end{equation}
Using the eigenstates in (\ref{eq:identity}) for $\mathbbm{1}^{(1)}$
and putting that result in (\ref{eq:rhoth_def}), we find the usual convex decomposition, 
\begin{equation}\label{eq:rho_conv}
\therm{(1)} = \sum_{n= 0}^{\infty}  \ket{n}{} p_n \bra{n}{}
\end{equation} 
with $\left|n\right\rangle =a_{n}^{\dagger} \vac $,
$p_{n}=\exp(-\beta E_{n})/Z^{(1)}$, and $Z^{(1)}=\sum_{n}\exp(-\beta E_{n})$.
Instead of actually constructing the wave functions $\phi_{n}(\mathbf{r})$
and their energies $E_{n}$ for a particular potential $v(\mathbf{r})$,
often one assumes that the  square potential $v(\mathbf{r})$ vanishes within
a very large cubical box of volume $V=L^{3}$ and is essentially infinite
outside it, confining the particle to the box; further, instead of
using the standing wave solutions of the resulting Hamiltonian, in
place of $\phi_{n}(\mathbf{r})$ one simply takes plane waves, $L^{-3/2}\exp(i\mathbf{k\cdot\mathbf{r})}$, satisfying periodic boundary conditions by requiring $k_{i}=2\pi n_{i}/L$
for each Cartesian component of $\mathbf{k}$, where the $n_{i}$
are integers \cite{CombescotShiauBook}. Then $n=\left\{ n_{x},n_{y},n_{z}\right\} $,
$E_{n}=\hbar^{2}(k_{x}^{2}+k_{y}^{2}+k_{z}^{2})/(2m)$, and in the
limit $V\rightarrow\infty$ the standard result is $Z^{(1)}\rightarrow V/\lambda^{3}$,
where $\lambda\equiv\sqrt{2\pi\hbar^{2}\beta/m}$ is the thermal de Broglie
wavelength. 
This is a good approximation for $\lambda^{3}/V\ll1$;
we present a new argument for this below, and show how corrections
to this limit could be included if desired. But note that whether
the $\phi_{n}(\mathbf{r})$ satisfying (\ref{eq:eigenvalues})
are adopted, or plane waves used in their stead, the wave functions
associated with the states appearing in the convex decomposition (\ref{eq:rho_conv}) are delocalized throughout the entire box confining the particles---as sketched in Fig.  \ref{fig:sketch}(b).

\section{Wave packets}\label{sec:method}
In contrast, we now introduce a strategy for rewriting
$\rho_{\rm th}^{(N)}$ that for $N=1$ will lead to a convex decomposition
of $\rho_{\rm th}^{(1)}$ in terms of wave packets, rather than completely
delocalized wave functions as in (\ref{eq:rho_conv}), and for arbitrary
$N$ will lead to a convex decomposition of $\rho_{\rm th}^{(N)}$ in
terms of sets of  wave packets. The strategy is based on finding an
operator $\mathcal{O^{\dagger}}(\mathbf{r})$ such that
\begin{equation}\label{eq:O_def}
e^{-\beta H/2}\psi^{\dagger}(\mathbf{r})=\mathcal{O^{\dagger}}(\mathbf{r})e^{-\beta H/2}.
\end{equation}
Once this is identified, we can ``pull through'' the factors $e^{-\beta H/2}$
appearing in the expression (\ref{eq:rhoth_def}) for the density operator
of the canonical ensemble until they act on the vacuum, which then
simply yields the vacuum state itself. The result is
\begin{align} \label{eq:th-general}
\begin{split}
\rho_{\rm th}^{(N)}={}&\frac{1}{N!}\frac{1}{Z^{(N)}}\int d\mathbf{r}_{1}d\mathbf{r}_{2}\cdots d\mathbf{r}_{N}\\
&\times\mathcal{O^{\dagger}}(\mathbf{r}_{1})\mathcal{O^{\dagger}}(\mathbf{r}_{2})\mathcal{\cdots O^{\dagger}}(\mathbf{r}_{N}) \vac  \\
&\times\cav \mathcal{O}(\mathbf{r}_{N})\cdots\mathcal{O}(\mathbf{r}_{2})\mathcal{O}(\mathbf{r}_{1}),
\end{split}
\end{align}
 a convex decomposition involving states that we shall see are $N$-particle
wave packets identified by 
$\mathcal{O^{\dagger}}(\mathbf{r}_{1})\mathcal{O^{\dagger}}(\mathbf{r}_{2})\mathcal{\cdots O^{\dagger}}(\mathbf{r}_{N})\vac $.

To determine $\mathcal{O^{\dagger}}(\mathbf{r})$ we introduce $\Psi^{\dagger}(\mathbf{r},\nu)\equiv e^{-\nu H/2}\psi^{\dagger}(\mathbf{r})e^{\nu H/2}$;
then $\Psi^{\dagger}(\mathbf{r},0)=\psi^{\dagger}(\mathbf{r})$ and
$\Psi^{\dagger}(\mathbf{r},\beta)=\mathcal{O^{\dagger}}(\mathbf{r})$.
We easily find 
\begin{equation} \label{eq:diff}
\begin{split}
\frac{\partial\Psi^{\dagger}(\mathbf{r},\nu)}{\partial\nu}&=\frac{1}{2}e^{-\nu H/2}\left[\psi^{\dagger}(\mathbf{r}),H\right]_{-}e^{\nu H/
2} \\
&=\frac{\hbar^{2}}{4m}\nabla^{2}\Psi^{\dagger}(\mathbf{r},\nu)-\frac{1}{2}v(\mathbf{r})\Psi^{\dagger}(\mathbf{r},\nu),
\end{split}
\end{equation}
where although only the commutator arises from the differentiation,
 the result holds whether we consider bosons
or fermions. 
This is a typical diffusion equation, and with the initial
condition at $\nu=0$, $\Psi^{\dagger}(\mathbf{r},0)=\psi^{\dagger}(\mathbf{r})$,
there are two special cases of interest: (a) if $v(\mathbf{r})$ is
neglected we find the solution 
\begin{equation}\label{eq:sol_diff}
 \Psi^{\dagger}(\mathbf{r},\beta)\underset{\underset{\rm everywhere}{v(\mathbf{r})\rightarrow 0}}\longrightarrow\int G(\mathbf{r}-\mathbf{r'};\lambda)\psi^{\dagger}(\mathbf{r'})d\mathbf{r'},
\end{equation}
with 
\begin{equation} \label{eq:G}
 G(\mathbf{r}-\mathbf{r'};\lambda)=\frac{2\sqrt{2}}{\lambda^{3}}e^{-2\pi\frac{\left|\mathbf{r}-\mathbf{r'}\right|^{2}}{\lambda^{2}}},
\end{equation}
where again $\lambda$ is the thermal de Broglie wavelength associated
with $\beta$, and $G(\mathbf{r}-\mathbf{r'};0)=\delta(\mathbf{r}-\mathbf{r'})$; 
 (b) if $v(\mathbf{r})$ is large and positive we have $\Psi^{\dagger}(\mathbf{r},\beta)\rightarrow0$
as $v(\mathbf{r})\rightarrow\infty$. Recalling from (\ref{eq:th-general}) that
$\mathcal{O^{\dagger}}(\mathbf{r})=\Psi^{\dagger}(\mathbf{r},\beta)$,
we see from (b) that if any of the $\mathbf{r}_{i}$ lie far outside
the nominal confining box, where $v(\mathbf{r})$ is large and positive,
the contribution from $\mathcal{O^{\dagger}}(\mathbf{r}_{i})$ can
be neglected. And since from (\ref{eq:G}) we see that $G(\mathbf{r}-\mathbf{r'};\lambda)$
is non-negligible for $\left|\mathbf{r}-\mathbf{r'}\right|$ less
than or on the order of $\lambda$, we can surmise from (\ref{eq:sol_diff}) that
the effect of $\mathcal{O^{\dagger}}(\mathbf{r}_{i})=\Psi^{\dagger}(\mathbf{r}_{i},\beta)$
can be described by (\ref{eq:sol_diff}) as long as $\mathbf{r}_{i}$ is within
a few $\lambda$ of any edge of the nominal box and inside it. Combining
these two results, if the volume $V$ of the nominal box satisfies
$V\gg\lambda^{3}$, we make a negligible error restricting the integrals
in (\ref{eq:th-general}) to the volume of the nominal confining box, 
and for all such points $\mathbf{r}$ within that volume, we  use 
\begin{equation}\label{eq:O_sol}
\mathcal{O^{\dagger}}(\mathbf{r})=\int G(\mathbf{r}-\mathbf{r'};\lambda)\psi^{\dagger}(\mathbf{r'})d\mathbf{r'}.\\
\end{equation}
Corrections  will be important as the confining
volume shrinks to a size on the order of the thermal de Broglie wavelength,
and indeed the approach we are developing here would be an interesting
way to explore such ``finite size'' effects. 
One would simply use the more correct (\ref{eq:th-general}), and instead of using (\ref{eq:G})
for the Green function use the exact solution of (\ref{eq:diff}). But we
defer such explorations to later communications, and here restrict
ourselves to a large volume and use  (\ref{eq:O_sol}) in (\ref{eq:th-general}) with the volume restricted to $V$.

It is convenient to introduce the normalized function
$\phi_{\mathbf{r},\lambda}(\mathbf{r')}\equiv\lambda^{3/2}G(\mathbf{r}-\mathbf{r'};\lambda),$
\begin{equation}\label{eq:wp}
 \phi_{\br,{\lambda}}(\br') = \frac{2\sqrt{2}}{\lambda^{3/2}} e^{-2 \pi \frac{|\br'-\br|^2}{\lambda^2}},
 \end{equation}
with  $ \int\left|\phi_{\mathbf{r},\lambda}(\br') \right|^{2}d\mathbf{r'}=1$.
Then, considering a large volume, we can write (\ref{eq:O_def})  as 
\begin{equation} \label{eq:rule}
 e^{-\beta H/2} \psi\dg(\br) = \frac{1}{\lambda^{3/2}} \left( \int \phi_{\br,{\lambda}}(\br') \psi\dg(\br') d\br' \right) e^{-\beta H/2},
\end{equation}
for any $\mathbf{r}$ within the volume $V$. Letting (\ref{eq:rule}) act on any state $\left|\xi\right\rangle $, multiplying
by the adjoint and integrating over all $\mathbf{r}$ within the volume $V$, we have
\begin{widetext}
\begin{equation}\label{eq:pullthrough} 
 \int_V d\br \left( e^{-\frac{\beta}{2} H} \psi\dg(\br) \ket{\xi}{} \bra{\xi}{} \psi(\br)  e^{-\frac{\beta}{2} H}\right) = \frac{V}{\lambda^3}\int_V \frac{d\br}{V} \left( \int d\br' \: \phi_{\br,\lambda}(\br') \psi\dg(\br') \right) e^{-\frac{\beta}{2} H} \ket{\xi}{} \bra{\xi}{}  e^{-\frac{\beta}{2} H}  \left( \int d\br'' \: \phi^*_{\br,\lambda}(\br'') \psi(\br'') \right) ,
\end{equation}
\end{widetext}
which is the central result of this section.

\section{A single particle} \label{sec:N1}
We now return to the example of a single particle, and use
(\ref{eq:pullthrough}) to write the expression for $\rho_{\rm th}^{(1)}$ that follows from (\ref{eq:th-general}) as 
\begin{equation} \label{eq:rho1_local1}
\rho_{\rm th}^{(1)}=\frac{1}{Z^{(1)}}\frac{V}{\lambda^{3}}\int_{V}\frac{d\mathbf{r}}{V}\left|\Phi_{\mathbf{r},\lambda}\right\rangle \left\langle \Phi_{\mathbf{r},\lambda}\right| ,
\end{equation}
where we have defined the single particle state 
 \begin{equation}\label{eq:wpN1_static}
 \ket{\Phi_{\br,\lambda }}{} \equiv \int d\br' \phi_{\br,\lambda }(\br') \psi\dg(\br') \vac.
\end{equation} 
Since the single particle states $\left|\Phi_{\mathbf{r},\lambda}\right\rangle $
are normalized, $\left\langle \Phi_{\mathbf{r},\lambda}|\Phi_{\mathbf{r},\lambda}\right\rangle =1$,
the condition $\Tr[\rho_{\rm th}^{(1)}]=1$ implies that 
\begin{equation} \label{eq:rho1_local}
\rho_{\rm th}^{(1)}=\int_{V}\frac{d\mathbf{r}}{V}\left|\Phi_{\mathbf{r},\lambda}\right\rangle \left\langle \Phi_{\mathbf{r},\lambda}\right|.
\end{equation}
Comparison with (\ref{eq:rho1_local1}) leads immediately to the result $Z^{(1)}=V/\lambda^{3}$, as would be expected.

Each single-particle state $\left|\Phi_{\mathbf{r},\lambda}\right\rangle $ has a coordinate
representation $\cav \psi(\mathbf{r'}) \left |\Phi_{\mathbf{r},\lambda}\right\rangle =\phi_{\mathbf{r},\lambda}( \br')$,
which is a minimum uncertainty wave packet centered at $\mathbf{r}$
and extending about a distance $\lambda$ from $\mathbf{r}$ in all
directions. In particular, for each of these states $\Delta x=\Delta y=\Delta z=\lambda/\sqrt{8\pi}$,
and $\Delta p_{x}=\Delta p_{y}=\Delta p_{z}=\sqrt{2\pi}\hbar/\lambda$.
Note that the expectation value of the momentum in each wave packet
$\phi_{\mathbf{r},\lambda}(\mathbf{r')}$ vanishes, but 
\begin{equation} \label{eq:Dp_tot}
\frac{\left(\Delta p_{x}\right)^{2}+\left(\Delta p_{y}\right)^{2}+\left(\Delta p_{z}\right)^{2}}{2m}=\frac{3}{2}k_{B}T.
\end{equation}
That is, in (\ref{eq:rho1_local}) we have the density operator written as a mixture
of localized wave packets, distributed in their mean positions over
the box containing the particle, but the expectation value of the
momentum of each wave packet vanishes, and the thermal energy is completely
contained in the width of the wave packet. A convex decomposition
more in line with the classical picture of thermal equilibrium would
involve localized wave packets with expectation values of momentum (Fig. \ref{fig:sketch}c),
but averaged over all possible momenta with an appropriate probability
distribution such that the average momentum of course vanishes.

Following a strategy introduced earlier \cite{Hornberger2003a}, we can move to such a convex decomposition by noting
that when (\ref{eq:wpN1_static}) is substituted into (\ref{eq:rho1_local}),
terms such as $\phi_{\mathbf{r},\lambda}(\mathbf{r'})\phi_{\mathbf{r},\lambda}^{*}(\mathbf{r}'')$
appear, which are then to be integrated over $\mathbf{r}$. That integral
extends over the volume $V$ of the box, but for the $\mathbf{r}'$
and $\mathbf{r}''$ of importance, it can be extended over
infinity without serious error, within our usual assumption of $V/\lambda^{3}\gg1$.
So for the $\mathbf{r}'$ and $\mathbf{r}''$ of importance we can write 
\begin{equation}\label{eq:trick}
\begin{split}
\int_{V}d\mathbf{r}&\phi_{\mathbf{r},\lambda}(\mathbf{r'})\phi_{\mathbf{r},\lambda}^{*}(\mathbf{r}'') \approx\int d\mathbf{r}\phi_{\mathbf{r},\lambda}(\mathbf{r'})\phi_{\mathbf{r},\lambda}^{*}(\mathbf{r}'')\\
&=e^{-\pi\frac{\left|\mathbf{r'}-\mathbf{r'}'\right|^{2}}{\lambda^{2}}}\\
  &=\int d\mathbf{r}\int d\mathbf{p}f_{\lambda_{m}}(\mathbf{p})\phi_{\mathbf{r}\mathbf{p},\lambda_{s}}(\mathbf{r'})\phi_{\mathbf{r}\mathbf{p},\lambda_{s}}^{*}(\mathbf{r}'')\\
 & \approx\int_{V}d\mathbf{r}\int d\mathbf{p}f_{\lambda_{m}}(\mathbf{p})\phi_{\mathbf{r}\mathbf{p},\lambda_{s}}(\mathbf{r'})\phi_{\mathbf{r}\mathbf{p},\lambda_{s}}^{*}(\mathbf{r}'').
 \end{split}
 \end{equation}
In the second strict equality we have introduced a normalized momentum distribution function, 
\begin{equation} \label{eq:f}
\begin{split}
 f_{\lambda_{m}}(\mathbf{p})= \left( \frac{ \lambda_m }{ 2 \pi \hbar} \right)^3 e^{-\lambda_m^2 \bp^2 / (4 \pi \hbar^2)} ,
\end{split}
\end{equation}
with $ \int f_{\lambda_{m}}(\mathbf{p})d\mathbf{p}=1$, and a set of normalized wave functions, 
\begin{equation} \label{eq:phi_gen}
\begin{split}
\phi_{\mathbf{r}\mathbf{p},\lambda_{s}}(\br')&=\frac{2\sqrt{2}}{\lambda_s^{3/2}} e^{i \bp\cdot (\br'-\br) / \hbar} e^{- 2\pi |\br'-\br|^2 / \lambda_s^2} ,
\end{split}
\end{equation}
with $\int\left|\phi_{\mathbf{r}\mathbf{p},\lambda_{s}}(\br') \right|^{2}d\mathbf{r'}=1$.
That strict equality holds as long as we choose $\lambda_{s}$
and $\lambda_{m}$ such that $\lambda^{-2}=\lambda_{s}^{-2}+\lambda_{m}^{-2}$.
Associating $\lambda_{s}$ and $\lambda_{m}$ with temperatures $T_{s}$
and $T_{m}$ respectively, through the relation of a thermal de Broglie
wavelength to temperature, this condition requires $T=T_{s}+T_{m}$.
We comment on the physical significance of $\lambda_{s}$ and $\lambda_{m}$
below, but we note that for $\mathbf{r}'$ and $\mathbf{r}''$ of
importance the last approximate result of (\ref{eq:trick}) follows if $\lambda_{s}$
is not too small, and that last result can then be applied for all
$\mathbf{r}'$ and $\mathbf{r}''$ within our usual assumption that
$V\gg\lambda^{3}$. 
Finally, inserting  (\ref{eq:wpN1_static}) in (\ref{eq:rho1_local}) and using  (\ref{eq:trick}), we find an alternate convex decomposition of $\rho_{\rm th}^{(1)}$,
\begin{equation}\label{eq:rho1_gen}
 \rho_{\rm th}^{(1)}=\int_{V}\frac{d\mathbf{r}}{V}\int d\mathbf{p}f_{\lambda_{m}}(\mathbf{p})\left|\Phi_{\mathbf{r}\mathbf{p},\lambda_{s}}\right\rangle \left\langle \Phi_{\mathbf{r}\mathbf{p},\lambda_{s}}\right|
\end{equation}
in terms of single particle states 
\begin{equation}\label{eq:wpN1} 
\ket{\Phi_{\br\bp,\lambda_s}}{} \equiv \int d\br' \phi_{\br \bp, \lambda_s}(\br') \psi\dg(\br') \vac.
\end{equation}
These are again minimum uncertainty states, but each has an expectation
value $\mathbf{p}$ of momentum, and $\Delta x=\Delta y=\Delta z=\lambda_{s}/\sqrt{8\pi}$,
and $\Delta p_{x}=\Delta p_{y}=\Delta p_{z}=\sqrt{2\pi}\hbar/\lambda_{s}$.
That is, these wave packets are broader than those introduced in (\ref{eq:rho1_local}). So instead of (\ref{eq:Dp_tot}), for these wave packets we have 
\begin{equation}\label{Dp_gen}
\begin{split}
 \frac{\left(\Delta p_{x}\right)^{2}+\left(\Delta p_{y}\right)^{2}+\left(\Delta p_{z}\right)^{2}}{2m}&=\frac{3}{2}k_{B}T_{s},\\
 \int\frac{p_{x}^{2}+p_{y}^{2}+p_{z}^{2}}{2 m }f_{\lambda_{m}}(\mathbf{p})d\mathbf{p}&=\frac{3}{2}k_{B}T_{m}.\\
\end{split}
\end{equation}
That is, part of the thermal energy resides in the motion of the wave packets, and part of it resides in their width.

For the purpose of illustration, consider argon atoms at room temperature ($\lambda = 0.16$\AA) \cite{Waber1965a}.
In the convex decomposition (\ref{eq:rho1_local}) involving the most localized wave
packets, the wave packet FWHM is $0.07\textrm{\AA}$.
Even in a convex decomposition (\ref{eq:rho1_gen}) that would involve wave packets
with only $1\%$ of the kinetic energy residing in the width of the
wave packets ($T_{s}=3K),$ and the remaining $99\%$ in the mean
velocities of the wave packets, the wave packet FWHM would be only
$0.75\textrm{\AA}$, which would be on the order of the diameter of
the outermost orbital of the argon electrons ($1.3\textrm{\AA}).$
Thus the construction of a convex decomposition of the form (\ref{eq:rho1_gen})
involving classical-like motion of wave packets is certainly possible
in the classical regime, as one would expect physically.

\section{Generalization to $N$ particles}\label{sec:N}
The generalization to a canonical ensemble of $N$ particles follows immediately,
applying (\ref{eq:rule}) to (\ref{eq:th-general}) $N$ times, ``pulling
through'' $\exp(-\beta H/2)$ until it acts on the vacuum state. We find 
\begin{align}\label{eq:rhoth_Nwp}
\begin{split}
\therm{(N)} ={}&  \frac{1}{N!} \frac{1}{Z^{(N)}} \left( \frac{V}{\lambda^3} \right)^N \int_V \frac{d\br_1 \dots d\br_N}{V^N}\\
&\times  \int d\bp_1 \dots d\bp_N f_{\lambda_m}(\bp_1) \dots f_{\lambda_m}(\bp_N)\\
&\times \ket{\Phi_{\br_1\bp_1\dots \br_N \bp_N, \lambda_s}}{}  \bra{\Phi_{\br_1\bp_1\dots \br_N \bp_N, \lambda_s}}{}, 
\end{split}
\end{align}
involving $N$-particle states with localized wave packets, 
\begin{align}\label{eq:wpN_dynamic}
 \ket{\Phi_{\br_1\bp_1\dots \br_N \bp_N, \lambda_s}}{} & \equiv {}  \left( \int d\br' \: \phi_{\br_1\bp_1,\lambda_s}(\br') \psi\dg(\br') \right) \times \dots  \nonumber \\
 \dots \times \Big( \int d\br'' &  \: \phi_{\br_N\bp_N,\lambda_s}(\br'') \psi\dg(\br'') \Big) \vac,
\end{align}
where $(\br_1, \dots, \br_N)$ and $(\bp_1, \dots, \bp_N)$ label the expectation values of the positions and momenta of the wave packets, respectively. 

While the thermal state density matrix has unit trace, $\Tr[\therm{(N)}]=1$, the individual states (\ref{eq:wpN_dynamic}) are not normalized, and their norm encompass all features distinguishing fermionic from bosonic statistics. This becomes obvious when we use the unit trace condition to identify the partition function,
\begin{align}
\begin{split}
Z^{(N)}={}&\frac{1}{N!}\left(\frac{V}{\lambda^{3}}\right)^{N} \int \frac{d\br_1 \dots d\br_N}{V^N}  \\
&\times \int d\bp_1 \dots d\bp_N f_{\lambda_m}(\bp_1) \dots f_{\lambda_m}(\bp_N) \\
&\times \left\langle{\Phi_{\br_1\bp_1\dots \br_N \bp_N, \lambda_s}} \vert {\Phi_{\br_1\bp_1\dots \br_N \bp_N, \lambda_s}}\right\rangle.
\end{split}
\end{align}
The factor $(1/N!)$ is the standard Gibbs factor \cite{HuangBook} that corrects the naive classical expression $(V/\lambda^{3})^{N}$ for the indistinguishability of the particles, and their product is
the usual result in the Maxwell-Boltzmann limit. The remaining integral
encapsulates the differences in the sum over states between bosonic
and fermionic systems.

\section{Illustrations} \label{sec:illu}
We now present some illustrations of our results. %
We begin with the partition function for a system of $N=2$ particles.
Of course, we recover the usual expression for the partition function
\begin{equation}
\begin{split}
Z^{(2)}  =  \frac{1}{2!} \left( \frac{V}{\lambda^3} \right)^2 \left( 1 \mp  \frac{\lambda^3}{V} \frac{1}{2\sqrt{2}} \right),
\end{split}
\end{equation}
which shows the quantum correction to the classical partition function (\textit{cf.} second term on the r.h.s.). 
This recovers the result for the first quantum corrections examined by Huang \cite{HuangBook}: the symmetry properties of the 2-particle wave function can be effectively accounted for with an attractive (repulsive) `statistical potential' for bosons (fermions), which depends on the temperature and therefore cannot be regarded as a true inter-particle potential. 
Note that this term vanishes in the Maxwell-Boltzmann limit ($\lambda \rightarrow 0$), and thus recovers the classical result. 
Importantly, our approach differs from that of Huang \cite{HuangBook} and other approaches conventionally presented in text books in that we consider the quantum state itself rather than the thermodynamics properties (derived from the partition function) to describe the ensemble.

\begin{figure*}[t!]
\includegraphics[width=2\columnwidth]{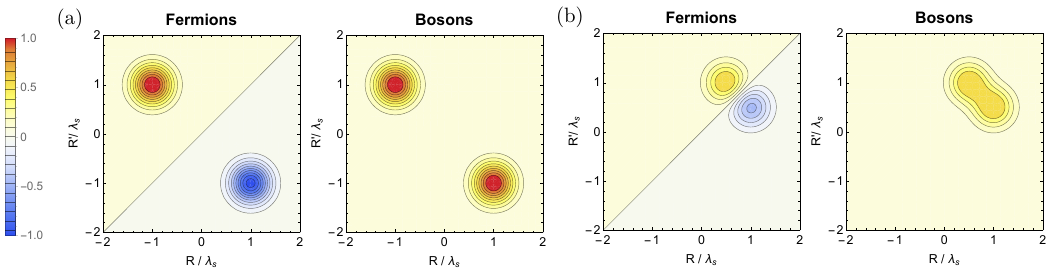}
\caption{Coordinate representations of a 2-particle wave packet (\ref{eq:coordN2}) for fermions and bosons with different relative centers. The average position of one wave-packet is fixed at $r_1/ \lambda_s=1$. (a) When the average positions of the wave packets are separated by a few $\lambda_s$ ($r_2 / \lambda_s= -1$)  the wave packets describe distinguishable particles, and the description thereby recovers the Maxwell-Boltzmann limit. (b) When the distance between the two centres is small, e.g. (b) $r_2 / \lambda_s =0.5$, the wave packets overlap, and we clearly see the anti-symmetry (symmetry) characterizing fermionic (bosonic) particles. Note that because the states (\ref{eq:wpN_dynamic}) are not normalized by construction the wave packets in (a) and (b) have different amplitudes. All vectors are taken to lie along a line (e.g. along the $\pm \hat{x}$ direction); all average momenta are zero here, i.e. $p_1=p_2=0$. }
\label{fig:coordN2}
\vspace*{-0.1cm}
\end{figure*}

As mentioned below Eq. (\ref{eq:wpN_dynamic}), the states introduced there are not normalized, and their norm depends on the nature of the particles: 
\begin{equation}
\left \langle \Phi_{\br_1 \br_2,\lambda} \vert \Phi_{\br_1 \br_2, \lambda} \right \rangle  = 1 \mp \, e^{-\frac{2\pi}{\lambda^2}|\br_1-\br_2|^2}. 
\end{equation}
The norm encompasses the different permutation rules for bosons or fermions which gives rise to different  quantum statistics. 
This can also be seen in Fig.  (\ref{fig:coordN2}), where we illustrate the coordinate representation of the  2-particle wave packet,
\begin{widetext}
\begin{equation}\label{eq:coordN2}
\frac{1}{\sqrt{2}}\cav \psi(\bR) \psi(\bR') \ket{\Phi_{\br_1\bp_1 \br_2 \bp_2, \lambda_s}}{}  =\frac{1}{\sqrt{2}} \Big(  \phi_{\br_1\bp_1,\lambda_s}(\bR') \phi_{\br_2\bp_2,\lambda_s}(\bR)  \mp \,  \phi_{\br_1\bp_1,\lambda_s}(\bR) \phi_{\br_2\bp_2,\lambda_s}(\bR') \Big),
\end{equation}
\end{widetext}
that is symmetric under exchange of coordinates for bosons, and anti-symmetric for fermions. 
If the density is low enough that for most $\br_1$ and $\br_2$ the wave packets appearing in the coordinate representation of the kets (and bras) in the convex decomposition (\ref{eq:rhoth_Nwp}) of $\therm{(2)}$ do not overlap, then to a good approximation the coordinate representations of the outer products of the kets and bras arising in (\ref{eq:rhoth_Nwp}) can be reorganized into terms involving simple products of two wave packets, as would be expected in the Maxwell-Boltzmann limit of distinguishable particles. This scenario generalizes to $\therm{(N)}$.

\begin{figure*}[t!]
\includegraphics[width=\textwidth]{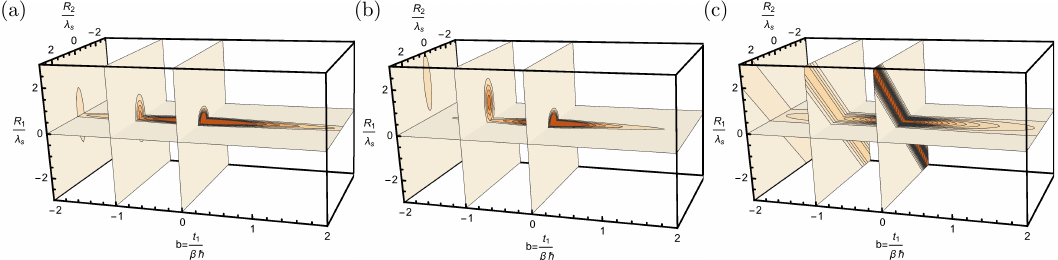}
\caption{Absolute value of the first-order correlation function (a;b) Eq.  (\ref{eq:G1N1}) for the single-particle state (\ref{eq:wpN1}) centred at $r=0$ with  $t_2=0$ and an average momentum of (a) $p=0$, (b)  $\lambda p/\hbar = 5$; (c) Eq. (\ref{eq:G1thN1}) for a one-particle thermal state, as a function of the measurement time $t_1$ and space points $\bR_1, \bR_2$.  
Comparing (a) with (b) shows how the wave-packet average momentum shifts the center of  the correlation function in the plane $(\bR_1, \bR_2)$ as a function of time, exhibiting the motion of the wave-packet center. Averaging each of the single-state contributions of either type (a) or type (b) according to (\ref{eq:G1_def}) recovers (c), the first-order correlation function of the complete one-particle thermal state (\ref{eq:G1thN1}), which only depends on the differences $\left|\bR_2 - \bR_{1}\right|$ and $(t_{2}-t_{1})$.
All vectors are taken to lie along a line. \label{fig:g1N1}}
\end{figure*}

We now turn to correlations functions, which
often enter in calculations involving thermal states. The $j^{\rm th}$
order correlation function for an $N$-particle thermal state is defined as 
\begin{widetext}
\begin{equation}\label{eq:Gth_def}
G^{(j)}_{\textrm{th}}(\bR_1 t_1, \dots, \bR_{j} t_{j}; \bR_{j+1} t_{j+1}, \dots, \bR_{2j} t_{2j}) \equiv \Tr^{(N)}\Big[\therm{(N)} \psi\dg(\bR_1 t_1) \dots \psi\dg(\bR_j t_j) \psi(\bR_{j+1} t_{j+1})  \dots \psi(\bR_{2j} t_{2j})\Big],
\end{equation}
\end{widetext}
where $\psi\dg(\bR t)$ is the Heisenberg representation of the creation operator. 
Since our convex decomposition (\ref{eq:rho1_gen}) involves localized wave packets,
the correlation functions can be written as integrals over correlation
functions associated with states consisting of sets of wave packets.
Consider first a single-particle system. From (\ref{eq:rho1_gen}), we find that the
first-order correlation function can be written as
\begin{equation}\label{eq:G1_def}
G^{(1)}_{\textrm{th}}(\bR_1 t_1; \bR_2 t_2)  \hspace{-0.05cm}= \hspace{-0.1cm} \int\hspace{-0.05cm} \frac{d\br}{V}\hspace{-0.08cm} \int d\bp f_{\lambda_m}(\bp)   G^{(1)}_{\br\bp,\lambda_s}(\bR_1 t_1; \bR_2 t_2),
\end{equation}
where the correlation function associated with the state $\left|\Phi_{\mathbf{r}\mathbf{p},\lambda_{s}}\right\rangle $ is
given by 
\begin{equation}\label{eq:G1N1}
G^{(1)}_{\br\bp,\lambda_s}(\bR_1 t_1; \bR_{2} t_{2}) \equiv  \bra{\Phi_{\br\bp,\lambda_s}}{} \psi\dg(\bR_1 t_1)  \psi(\bR_{2} t_{2}) \ket{\Phi_{\br\bp,\lambda_s}}{}.
\end{equation}
Each of these state correlation functions can be written as
$G^{(1)}_{\br\bp,\lambda_s}(\bR_1 t_1; \bR_{2} t_{2}) = \mathscr{F}_{\br \bp,\lambda_s}(\bR_1t_1)  \mathscr{F}^*_{\br \bp,\lambda_s}(\bR_2 t_2),$
where we have introduced the function
\begin{align}
\mathscr{F}_{\br \bp,\lambda_s}(\bR t) \equiv {}&\left( \frac{2}{\lambda(1+2 i b)}\right)^{\frac{3}{2}}  e^{-\frac{1}{8\pi}\left( \frac{\lambda_s \bp}{\hbar}\right)^2}\nonumber \\
&\times e^{{-\frac{2\pi}{1+2 i b} \left( \frac{\br-\bR}{\lambda_s} - \frac{i}{4\pi} \frac{\lambda_s \bp}{\hbar}\right)^2}},
\end{align}
with $b\equiv t / (\beta \hbar)$. Fig. \ref{fig:g1N1} illustrates the first-order correlation function (\ref{eq:G1N1}) for (a) static wave packets (zero average momentum), and (b) dynamic wave packets. 

Since the individual expectation values (\ref{eq:G1N1}) involve a single
wave packet, each of these wave packet contributions are products of functions
of $(\mathbf{R}_{1},t_{1})$ and $(\mathbf{R}_{2},t_{2})$, but the
integral over all these contributions (\ref{eq:G1_def}) gives a correlation function
\begin{align}\label{eq:G1thN1}
G^{(1)}_{\textrm{th}}(\bR_1 t_1; \bR_2 t_2) =  {}&\frac{1}{V} \left( 1 + i \frac{t_2 - t_1}{\beta \hbar}\right)^{-\frac{3}{2}}\\
&\times e^{{-\frac{\pi}{1+ i \frac{t_2 - t_1}{\beta \hbar} }  \left|\frac{\bR_2 - \bR_1}{ \lambda}\right|^2}} \nonumber 
\end{align}
 that depends only on $\left|\bR_2 - \bR_{1}\right|$ and $(t_{2}-t_{1})$, as illustrated in Fig. \ref{fig:g1N1}c.

\begin{figure}
\includegraphics[width=0.5\textwidth]{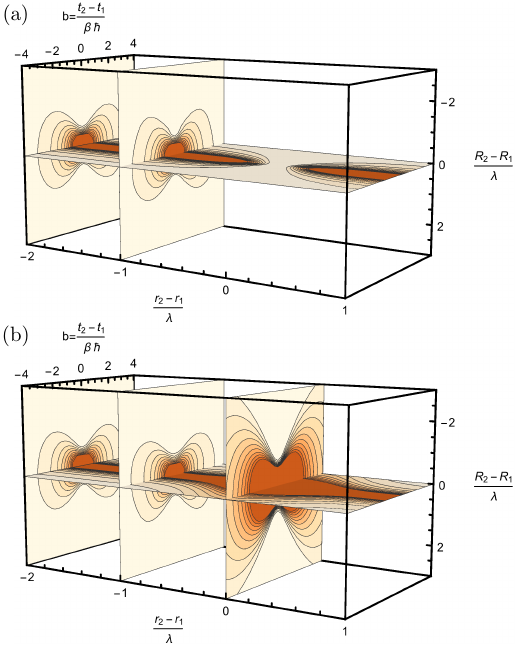}
\vspace*{-5ex}
\caption{First-order correlation function  (\ref{eq:G1N2}) for the 2-particle states for (a) fermions and (b) bosons. We clearly see the consequence of Pauli exclusion principle and the anti-bunching character of fermions, leading to a zero correlation function for overlapping wave packets (${|\br_2-\br_1|}/{\lambda} = 0$). In contrast, the bosons (b) present an enhanced correlation function due to their bunching character. For non-overlapping wave functions (${|\br_2-\br_1|}/{\lambda}\gg 1$), both fermionic and bosonic systems exhibit the same properties and we recover the Maxwell-Boltzmann limit, i.e. a profile similar to Fig. (\ref{fig:g1N1}a) for individual particles. All vectors are taken to lie along a line (e.g. along the $\pm \hat{x}$ direction).
\label{fig:G1N2}}
\vspace*{-1em}
\end{figure}

We now turn to a system with $N=2$ particles. If we consider the first order correlation function we
will find contributions from each state in (\ref{eq:rhoth_Nwp}), with each contribution
arising from a first order correlation function of a state with two
wave packets. Here the nature of the particles will appear explicitly.
For example, if we consider a state involving two wave packets with
no average momentum, $p_{1}=p_{2}=0$, we find 
\begin{widetext}
\begin{align}\label{eq:G1N2}
\begin{split}
G^{(1)}_{\br_1 0 \br_2 0,\lambda_s}(\bR_1 t_1; \bR_{2} t_{2}) \equiv {}& \bra{\Phi_{\br_1 0 \br_2 0 ,\lambda_s}}{} \Psi\dg(\bR_1 t_1)\ \Psi(\bR_{2} t_{2}) \ket{\Phi_{\br_1 0 \br_2 0,\lambda_s}}{} \\
={}& G^{(1)}_{\br_10,\lambda_s}(\bR_1 t_1; \bR_{2} t_{2}) +  G^{(1)}_{\br_20,\lambda_s}(\bR_1 t_1; \bR_{2} t_{2}) \\
&\mp e^{-\pi\frac{|\br_1-\br_2|^2}{\lambda_s^2}} \Big( \mathscr{F}_{\br_1 0,\lambda_s}(\bR_1 t_1) \mathscr{F}_{\br_2 0,\lambda_s}^*(\bR_2 t_2) + \mathscr{F}_{\br_2 0, \lambda_s}(\bR_1 t_1)\mathscr{F}^*_{\br_1 0,\lambda_s}(\bR_2 t_2)\Big), 
\end{split}
\end{align}
\end{widetext}
where note that the contribution involves a single-particle term from
each wave packet, plus quantum corrections that appear for overlapping
wave-functions (see Fig. \ref{fig:G1N2}). In the Maxwell-Boltzmann limit the contribution from
these correction terms will become negligible, as indeed they will
be for an $N$ particle system, and the first order correlation function
then reduces to the contributions from the individual wave packets,
\begin{equation}\label{eq:GN_MB}
G^{(1)}_{\br_1 \dots \br_N 0\dots 0,\lambda}(\bR_1 t_1; \bR_{2} t_{2}) \longrightarrow \sum_{n=1}^{N} G^{(1)}_{\br_n 0 ,\lambda}(\bR_1 t_1; \bR_{2} t_{2}).
\end{equation}

\section{Outlook and Conclusions}

We have shown that an $N$-particle thermal state can be represented by a mixture of  $N$-particle localized wave-packets, and have given expressions for such decompositions. From these decompositions, one can see explicitly how the classical particle picture arises as a limit of the quantum wavepackets. 

Previous work \cite{Chenu2015c} on the density operator for light in thermal equilibrium showed that a convex decomposition in terms of single pulses is not sufficient, and hinted that perhaps decompositions involving sets of pulses would be appropriate.  Here we have shown that an analogous situation arises for non relativistic particles in thermal equilibrium, where it is perhaps more obvious.  A convex decomposition into states involving only a single particle obviously cannot describe higher order correlation functions correctly. 
 To reproduce second-order functions in an
$N$-particle system, one would need a (improperly normalized) mixture
of at least two-particle states.  More generally, reproducing the $n$-th order correlation function in an $N$-particle system with a normalized mixture would require $N$-particle states.  With this extension, \emph{any} order correlation function can be reproduced. 

We point out that the convex decompositions
of thermal equilibrium we have presented here can be used to generate
even more decompositions. Since the density operator commutes with the Hamiltonian,
we clearly have $\rho_{\rm th}^{(N)}=e^{-iHt/\hbar}\rho_{\rm th}^{(N)}e^{iHt/\hbar}$.
Taking $H$ to be the free particle Hamiltonian, forming $e^{-iHt/\hbar}\rho_{\rm th}^{(N)}e^{iHt/\hbar}$
from (\ref{eq:rhoth_Nwp}) it is clear that the many-particle wave functions associated
with the kets $e^{-iHt/\hbar}\left|\Phi_{\mathbf{r}_{1}\mathbf{p}_{1}\cdots\mathbf{r}_{N}\mathbf{p}_{N}}\right\rangle $
will involve superpositions of products of wave packets that have
spread from their initial minimum uncertainty states to states with
broader widths, for either $t$ positive or negative; for $\left|t\right|$
not too large the use of the free particle Hamiltonian will not introduce
significant error in our limit $V\gg\lambda^{3}$ of interest. Thus
the decompositions we have detailed up until this point, which have
all involved minimum uncertainty wave packets, are only a small subset
of those possible.

We also note that the approach
developed here extends immediately to the grand canonical ensemble.
The density operator for that ensemble is given by 
\begin{equation}
\rho_{\rm th}^{\rm GC}=\frac{e^{-\beta H-\mu\mathcal{N}}}{\Tr\left[e^{-\beta H-\mu\mathcal{N}}\right]},
\end{equation}
where $\mu$ is the chemical potential and $\mathcal{N}$ the number
operator. In the usual way this can be written as 
\begin{equation}
\rho_{\rm th}^{\rm GC}=\sum_{N}\frac{e^{-\mu {N}}Z^{(N)}}{\Tr\left[e^{-\beta H-\mu \mathcal{N}}\right]}\rho_{\rm th}^{(N)},
\end{equation}
where we take $\rho_{\rm th}^{(N)}$ to be given in the form (2). Then
convex decompositions for all the $\rho_{\rm th}^{(N)}$ can be constructed
as we have discussed, and a convex decomposition for $\rho_{\rm th}^{\rm GC}$ involving
sets of wave packets with different numbers in different subsets can
be constructed.

In conclusion, we have constructed new convex decompositions of the density operator for nonrelativistic particles in thermal equilibrium. 
This manuscript focuses on presenting the method to develope the formalism, 
and is limited to non-interacting particles. 
 We note that an alternative derivation has been recently proposed  \cite{Chenu2017c}.
The many-particle
wave functions associated with the kets in the decompositions involve
sets of wave packets, in general with a range of average positions
and momenta. The combination of amplitudes in the many-particle wave
functions capture the bosonic or fermionic character of the particles,
and the distributions of positions and momenta of the wave packets
allow for a connection with the usual classical picture of thermal
equilibrium. 
 This representation explicitly contains the two essential features proposed as a requirement for thermal states, namely stochasticity and  spatial extension of the particle wave function \cite{Drossel2017a}.  

 In the Maxwell-Boltzmann limit this kind of decomposition
has already proven useful in decoherence calculations \cite{Hornberger2003a},
allowing the effect of an environment to be obtained from scattering calculations with single particles. 
Extensions to fermions and bosons presented here should allow for similar simplifications. 
\\
\emph{Acknowledgments.---}  We are grateful to M. Combescot for insightful discussions. AC and AMB thank the Kavli Institute for Theoretical Physics for hosting them during the Many-Body Physics with Light program. AC thanks P. Brumer and J. Cao for hosting her during the completion of this work. This research was supported in part by the Swiss National Science Foundation (AC), the Natural Sciences and Engineering Research Council of Canada (JES), the National Science Foundation under Grant No. NSF PHY11-25915 (KITP), and by Perimeter Institute for Theoretical Physics, which includes support from the Government of Canada through the Department of Innovation, Science and Economic Development Canada and by the Province of Ontario through the Ministry of Research, Innovation and Science.

%\bibliography{main}
%\bibliographystyle{apsrev4-1}

%

\end{document}